\documentclass[aps,pre,reprint,10pt,superscriptaddress,showpacs]{revtex4-1}
\usepackage{amsfonts} 
\usepackage{amsmath}
\usepackage{amssymb}
\usepackage{graphicx}
\usepackage{color}
\usepackage{caption}
\begin{document}
\title{Generation of wakefields  and  electromagnetic solitons  in relativistic degenerate plasmas}
\author{Sima Roy }
\affiliation{ Department of Mathematics, Siksha Bhavana, Visva-Bharati (A Central University), Santiniketan-731 235, West Bengal, India}
\author{Debjani Chatterjee}
\affiliation{ Department of Mathematics, Siksha Bhavana, Visva-Bharati (A Central University), Santiniketan-731 235, West Bengal, India}
\affiliation{ Ahmadpur Joydurga Girls' High School, Birbhum, West Bengal, India}
\author{A. P. Misra}
\email{apmisra@visva-bharati.ac.in, apmisra@gmail.com}
\affiliation{ Department of Mathematics, Siksha Bhavana, Visva-Bharati (A Central University), Santiniketan-731 235, West Bengal, India}



\begin{abstract}
The nonlinear interaction of intense linearly polarized electromagnetic  waves (EMWs) with longitudinal electron density perturbations is revisited in relativistic degenerate plasmas.  The nonlinear dynamics of the EMWs and the longitudinal field, driven by the EMW ponderomotive force, is  governed by a coupled set of nonlinear partial differential equations.     A numerical simulation of these coupled equations reveals that the generation of wakefields is possible in  weakly relativistic degenerate plasmas with  $R_0\equiv p_F/mc\ll1$ and $v_g/c\sim1$, where $p_F$ is the Fermi momentum, $m$ is the mass of electrons, $c$ is the speed of light in vacuum, and $v_g$ is the EMW group velocity. However, when the ratio   $v_g/c$ is reduced to $\sim0.1$, the wakefield generation is suppressed, instead the longitudinal fields get localized to form soliton-like structures. On the other hand,   in the regimes of moderate $(R_0\lesssim1)$  or strong  relativistic degeneracy   $(R_0>1)$  with $v_g/c\sim0.1$,    only the   EM solitons can be formed.     
\end{abstract}
\maketitle
\section{Introduction \label{sec-intro}} 
The theory of generation of wakefields  has reached a significant milestone in plasmas since   it was first introduced by Tajima and Dawson \cite{tajima1979} more than three decades ago. This mechanism has been  useful  for not only its fundamental importance, but also its applications to particle acceleration schemes. When  an intense electromagnetic (EM)   pulse propagates in a plasma, it  induces a longitudinal plasma oscillations (wake) behind the pulse driven by the EM wave  ponderomotive force.  Electrons are then trapped into the wake and   accelerated to extremely high energies (GeV-TeV).   Recently, a large interest has been devoted to the generation of  wakefields in plasmas both theoretically \cite{gorbunov1987,balakirev2000,malka2008,martins2010,lu2007,mironov1992,chen2002,chen2009,misra2010,brodin1998,shukla1999,shukla2009} and experimentally \cite{kneip2009,kim2013,leemans2014}, where  a laser pulse  traveling close to the speed of light in vacuum $(c)$ is considered as an effective source for particle acceleration.   In this context, a number of alternative schemes including the use of proton bunches for driving plasma wakefield accelerators has also been proposed (See, e.g., Refs.  \cite{shen2007,najmudin2003,trines2009,shukla2009,joshi2018,litos2016}).    
\par
Previous investigations \cite{misra2010,brodin1998,shukla2009,bulanov1999} indicate that besides the wakefield generation,  various other nonlinear phenomena including the formation of coherent structures can be observed due to the nonlinear interaction of intense EM waves (EMWs) with plasmas.      If the pulse size is long compared to the skin depth, wakefield generation is not so pronounced. The plasma number density also plays an important role in the transition from wakefield generation to soliton formation \cite{amol2018}. It has been reported by means of one-dimensional (1D)  particle-in-cell (PIC) simulation  that  whenever plasma density approaches a critical value, EM pulse undergoes nonlinear self-interaction and eventually  soliton-like structures are formed \cite{amol2018}.   In Ref. \cite{bulanov1999}, the theory of the formation of solitons   in  underdense plasmas has been studied using 1D PIC simulation.  Nonlinear 1D relativistic solitons were studied analytically in Refs. \cite{mima1986,kaw1992,mikaberidze2015} and numerically in Refs. \cite{esirkepov1998,bulanov1992}. The formation of EM solitons and their properties have also been studied in a number of other works  \cite{shukla2005, misra2010a,saxena2013,siminos2014}.   It has been shown in Refs. \cite{mamun2012,roy2012} that the the relativistic degeneracy of  electrons can significantly modify the dispersion properties of EM waves and profiles of nonlinear coherent structures. However, in these works   the authors have  considered the degeneracy of electrons   in two particular limits: weakly relativistic and ultra-relativistic degeneracy.    While these works and the present work have   similarities at some particular   considerations of two relativistic degeneracy limits, however, our theory is applicable to arbitrary degeneracy of electrons as well. Furthermore, the aim of the present work is to present   some theoretical results for the generation of wakefields and their transition to EM solitons, as well as the formation of EM solitons which are distinctive to the results of  Refs. \cite{mamun2012,roy2012}.
\par
In this paper, we consider the excitation of wakefields, their transition from the wakefield generation to the formation of soliton-like structures, as well as the formation of EM solitons using the same relativistic fluid model as in Ref. \cite{misra2018}.   Our starting point is the nonlinear coupled equations for the  EMW field and the density perturbations of low-frequency electron plasma oscillations that are reinforced by the EMW's pondermotive force.  We solve these equations numerically to study the competing mechanisms between the wakefield generation and the localization, as well as the formation of EM solitons. It is shown that both the relativistic degeneracy parameter $R_0$ and the group velocity $v_g$ of EMW play  crucial roles for the generation of wakefields  and the formation of  solitary pulses. 
\section{Relativistic fluid model and evolution equations \label{sec-fluid-model}}
Our basic assumption is that  plasma comprises relativistic  degenerate electrons and stationary ions, and  plasma interacts nonlinearly with intense EMWs.   We also assume that the finite amplitude linearly polarized EMW propagates in the $z$-direction, i.e., all the field variables are functions of $z$ and $t$. In the  dynamics of relativistic degenerate electrons, we include the weakly relativistic effects  on the particle motion in the EMW fields, but fully relativistic effects on the particle thermal motion.    Thus, the  $z$-components of the relativistic 1D fluid equations are \cite{misra2018} 
 \begin{equation}
 \frac{d}{dt}\left(\frac{\gamma H v_{z}}{nc^2}\right)=e\left(\frac{\partial\phi}{\partial z}-\frac{1}{2}\frac{e n}{\gamma H}\frac{\partial A_x^2}{\partial z}\right)-\frac{1}{n_l}\frac{\partial P}{\partial z}, \label{moment-eq-parallel}
 \end{equation}
  \begin{equation}
 \frac{\partial n_l}{\partial t} +\frac{\partial}{\partial z} (n_l {v}_{z})=0,\label{continuity-eq-reduced}
 \end{equation}
 \begin{equation}
 \frac{\partial^2 \phi}{\partial z^2}=4\pi e \left(n_l-n_0\right), \label{poisson-eq-reduced}
 \end{equation}
 and the transverse or $x$-component of the EMW equation, given by,
 \begin{equation}
  \frac{\partial^2{\bf A}}{\partial t^2}-c^2\nabla^2{\bf A}+c\frac{\partial}{\partial t}(\nabla \phi)+4\pi e cn_l{\bf v}=0, \label{em-wave-eq}
  \end{equation}
   is
 \begin{equation}
\left(\frac{\partial^2}{\partial z^2}- \frac{1}{c^2}\frac{\partial^2}{\partial t^2}\right) {A}_x=\frac{4\pi e}{c} \gamma nv_{x}, \label{em-wave-transv}  
\end{equation}
where $n$ is the electron number density with $n_0$ denoting its equilibrium value (i.e., the constant density of the neutralizing background) and $n_l\equiv n\gamma$ the density in laboratory frame,  $e$ is the elementary charge, $m$ is the electron mass, $v_z$ is the parallel component of the electron velocity ${\bf v}$, $c$ is the speed of light in vacuum,   $H={\cal E}+P$ is the enthalpy per unit volume measured in the rest frame of each element of the fluid in which     ${\cal E}=mnc^2+\bar{\epsilon}$ is the total energy density  with $\bar{\epsilon}$ denoting the  internal energy of the fluid and   $P$ is the degeneracy pressure of electrons.  Also, $d/dt\equiv\partial/\partial t+ v_{z} \partial/\partial z$,  ${\bf A}=\left(A_x(z,t),0,0\right)$ is the vector potential and $\phi(z,t)$ is the scalar potential given by   the transverse and longitudinal components of  the electric field, i.e., $E_x(z,t)=-{\partial A_x(z,t)}/{c \partial t}$ and   $E_z(z,t)=-\partial\phi(z,t)/\partial z$. Furthermore,   $v_{x}=A_x{e nc}/{\gamma H}$ is the quiver velocity and $\gamma$ is the relativistic factor, given by, \cite{misra2018}
 \begin{equation}
 \gamma=\sqrt{\frac{1+ e^2n^2  A^2_x/H^2}{1-v_{z}^2/c^2}}. \label{gamma}
\end{equation}
The relativistic degeneracy pressure $P$ and the  total energy density  ${\cal E}$  of electron fluids are given by \cite{chandrasekhar1935}
\begin{equation}
 \begin{split}
 &\left(P,{\cal E}\right)=\frac{m_e^4c^5}{3\pi^2\hbar^3}\left[f(R),~R^3\left(1+R^2\right)^{1/2}-f(R)\right],\\
 &f(R)=\frac18\left[ R\left(2R^2-3\right)\left(1+R^2\right)^{1/2}+3\sinh^{-1}R\right],
 \end{split} \label{pressure}
\end{equation}  
where $\hbar=h/2\pi$ is the reduced Planck's constant, $R=p_{F}/mc=\hbar\left(3\pi^2n\right)^{1/3}/mc$ is the dimensionless degeneracy parameter, and $H\equiv P+{\cal E}=nmc^2\sqrt{1+R^2}$.
\par
Equations \eqref{moment-eq-parallel} to \eqref{poisson-eq-reduced} constitute the basic equations for the description of   low-frequency electron plasma oscillations that are driven by the EMW pondermotive force and these equations are coupled to the EMW  equation  \eqref{em-wave-transv}. Next, we derive a  set of reduced equations from Eqs. \eqref{moment-eq-parallel} to \eqref{poisson-eq-reduced} and \eqref{em-wave-transv} for the slow motion approximation of relativistic dynamics of electrons, i.e., when   $\gamma \approx 1+(1/2)\left({\bf v}/c\right)^2$.   Here,  we note that instead of considering the longitudinal component of the EM wave equation \eqref{em-wave-longi}, i.e., 
\begin{equation}
\frac{\partial^2\phi}{\partial t\partial z}=-4\pi en_l v_{z}.\label{em-wave-longi}
\end{equation}
we consider the Poisson equation \eqref{poisson-eq-reduced}, because  the electrostatic potential $\phi$, associated with the electric field $E_z$, is created due to the density variations of charged particles.   In fact, Eq. \eqref{em-wave-longi}  is consistent with the estimates for the density perturbations $n_1~(n=n_0+n_1,~n_1\ll n_0)$ and potential $\phi$.   We also assume that $en A_x/H  \sim o(\epsilon)$ for which  the quiver velocity, $v_{x}/c\sim o(\epsilon)$, and   Eqs. \eqref{moment-eq-parallel} and \eqref{em-wave-longi} yield $\phi,~n_{1},~v_{z}\sim o(\epsilon^2)$. Next, linearizing   Eqs. \eqref{moment-eq-parallel} to \eqref{poisson-eq-reduced} and \eqref{em-wave-transv} about the equilibrium state, i.e., $n_l\equiv n\gamma=n_0+N,~H=H_0+H_1$ etc.,  and normalizing the physical quantities according to $t\rightarrow t\sqrt{\eta}\omega_p$, $z\rightarrow z\sqrt{\eta}\omega_p/c$, $A\equiv \eta eA_x/mc^2$, where   $\eta=1/\sqrt{1+R_0^2}$ and $\omega_p=\sqrt{4\pi n_0e^2/m}$ (For details, readers are referred to Ref. \cite{misra2018}), we obtain the following coupled equations for low-frequency electron density perturbations that are driven by the EM wave ponderomotive force and the EM wave field \cite{misra2018}.
\begin{equation}
\left(\frac{\partial^2}{\partial t^2}-\delta\frac{\partial^2}{\partial z^2}+1\right)N=\frac{1}{2} (1-\delta)\frac{\partial^2  |A|^2}{\partial z^2}, \label{electron-main-eq}
\end{equation} 
\begin{equation}
\left(\frac{\partial^2}{\partial t^2}- \frac{\partial^2}{\partial z^2}+1\right)A + (1-\delta)(N-\alpha  |A|^2)A=0,\label{cpem-main-eq}
\end{equation}
where $\alpha=\eta^2/2$, $\delta=(1-\eta^2)/3$ and $\eta=1/\sqrt{1+R_0^2}$ and  $R_0=\hbar\left(3\pi^2n_0\right)^{1/3}/mc$ is a measure of the strength of plasma degeneracy, i.e., $R_0\ll1$ corresponds to weakly relativistic degenerate plasma, whereas  $R_0\gg1$ is refereed to as ultra-relativistic degenerate  plasmas.     
\section{Results \label{sec-results}}
 In this section we study the excitation of   wakefields, their  transition   into the formation of localized structures, as well as the formation of EM  solitons  by solving Eqs. \eqref{electron-main-eq} and \eqref{cpem-main-eq} numerically. The results will be presented in subsections \ref{sec-sub-wakefield} to \ref{sec-sub-em-soliton}.  We look for solutions in a moving frame of reference, given by, $\xi=z-v_gt$ and $\tau=t-v_gz$, where   $v_g$ stands for the  group velocity and $1/v_g$ the phase velocity of the EMW respectively. We assume that  the vector potential $A$ is of the form $A=a(\xi) \exp(-i\omega \tau)$ and $N$ depends only on the variable $\xi$. Here, $a$ is a real function of $\xi$ and    $\omega$  the wave frequency.      Thus, Eqs. \eqref{electron-main-eq} and \eqref{cpem-main-eq} reduce  to
 \begin{equation}
(v_g^2-\delta)\frac{d^2N}{d\xi^2}+N=\left(1-\delta\right)\left(\frac{da}{d\xi}\right)^2+a(1-\delta)\frac{d^2a}{d\xi^2}, \label{reduced-electron-main-eq}
\end{equation}
\begin{equation}
\frac{d^2a}{d\xi^2}+\left[ \omega^2-\frac{1+\left(1-\delta\right)\left(N-\alpha a^2\right)}{1-v_g^2} \right] a=0. \label{reduced-cpem-main-eq}
\end{equation}
It is not an easy task to find an analytic  solution of Eqs. \eqref{reduced-electron-main-eq} and \eqref{reduced-cpem-main-eq}, however,  we can numerically solve these equations by Newton's method with the boundary conditions $N,~a,{dN}/{d\xi}$ and ${da}/{d\xi}\rightarrow 0$ as $\xi\rightarrow \pm \infty$.  We note that these equations are reversible under the transformation $\xi\rightarrow -\xi,~ N\rightarrow N$ and $a\rightarrow \pm a$. So, one can look for the symmetric solutions for $N$, while either symmetric or antisymmetric for $a$.
  \subsection{Generation of wakefields \label{sec-sub-wakefield}}
We consider the excitation of wakefields by the EM wave driven ponderomotive force.   For a Gaussian driving pulse of the form $a\sim a_0\exp\left(-\xi^2/L_p^2\right)$, where $L_p$ is the size of the driving pulse, the driver is resonant for $k_pL_p=\sqrt{2}$, and in units of time $\sqrt{2}/\omega_p$. For the generation of wakefield, the driving pulse size must satisfy $L_p<\lambda_p\sim c/\omega_p$, the typical wavelength of plasma oscillation. 
However,    the dispersive  effect $(\sim \delta \partial^2/\partial z^2)$ in Eq.  \eqref{electron-main-eq}  can  forbid   the generation of wakefields due to some finite values of the degeneracy parameter $R_0$. Typically, the parameter $\delta$ involves the two dispersive effects due to (i)  the charge  separation of plasma particles (Poisson equation) and (ii) the gradient of relativistic degeneracy pressure $(\propto R_0)$. The latter with  $R_0=0$ corresponds to the classical cold plasma limit where the dispersive term $(\propto\delta)$ in Eq. \eqref{electron-main-eq} disappears. So, values of $R_0\ll1$ for which $\delta\ll1$  may favor the generation of wakefields. In a frame moving with the constant group velocity $v_g$ of the EM pulse,   the similar phenomenon can also occur, i.e.,   the generation of wakefields is  also possible for $v_g\sim c$ and  $R_0\ll1$  (or $\delta\ll1$), i.e., in the weakly relativistic degenerate plasmas.
\par
\begin{figure*}
\includegraphics[scale=0.5]{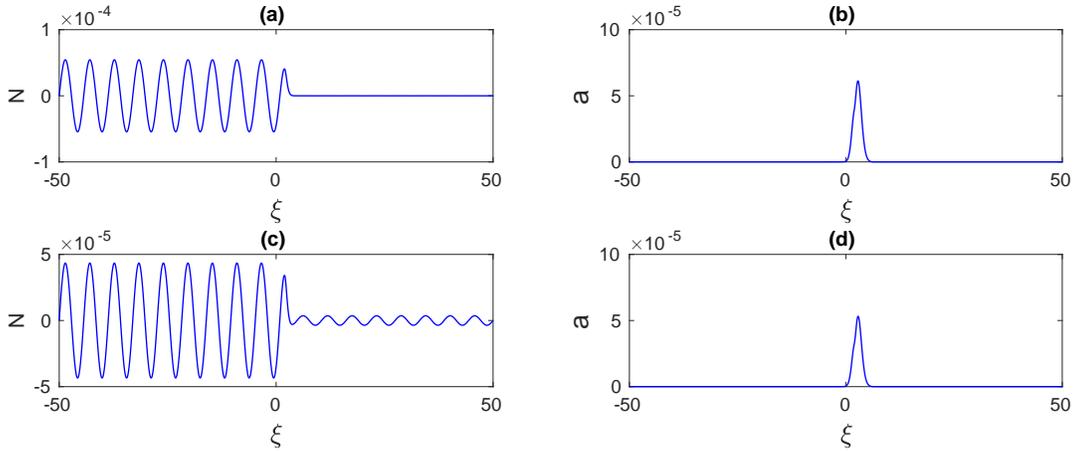}
\caption{ \label{fig1} Plasma wakefields [numerical solution of Eqs. \eqref{reduced-electron-main-eq} and \eqref{reduced-cpem-main-eq}] driven by the EM wave's pondermotive force are shown for different values of the relativistic degeneracy parameter $R_0$:  $R_0 =0.005$ (upper panel) and $R_0=0.05$ (lower panel) with   fixed values of $\omega=0.78$, $v_g=0.9$,   $L_p=0.1$ and $a_0=0.01$. The wakefield generation is  suppressed for $R_0>0.01$, i.e.,   the structure of the wakefield is retained for $0\lesssim R_0 \lesssim 0.01,~v_g=0.9,~L_p=0.1$, and $0.67\lesssim\omega<1$. The generation of wakefield also occurs in some other parameter regimes with $0.01\lesssim R_0\lesssim0.014,~v_g=0.43,~L_p=0.1$, and $ \omega=0.82$ (not shown in the figure).}
\end{figure*}

 
 Figure \ref{fig1} shows the profiles of the longitudinal oscillations of the electron density perturbation [subplots (a) and (c)] and the perturbed EM wave field  [subplots (b) and (d)]. It is seen that the  electron density perturbation has a pure weakfield nature [subplot (a)], i.e., after the peak of the EM pulse [subplot (b)], harmonic oscillations are generated with pick amplitude of the pulse $\sim10^{-5}$. The  generation of such   wakefields  occurs   at a lower amplitude of the driving pulse, i.e.,  $a_0\sim0.01$ together with the group velocity close to $c$, i.e., $v_g\sim0.9$ (in dimensionless), and with a lower value of the degeneracy parameter $R_0 =0.005$ which corresponds to a regime with  $n_0=7.33\times10^{28}$ m$^{-3}$, and the structure of the wakefield is retained for $R_0\sim0.01$ and $0.67\lesssim\omega<1$ as well for a fixed pulse size $L_p=0.1$. 
 \par 
The generation of  wakefields also occurs in some other parameter regimes with $0.01\lesssim R_0\lesssim0.014,~v_g=0.43,~L_p=0.1$, and $ \omega=0.82$.  However, as the value of $R_0$ increases from $R_0=0.005$ to $R_0=0.05$, i.e., one enters from relatively low-to high-density  regimes with $n_0=7.33\times10^{31}$ m$^{-3}$  or the value of $v_g$ is further lowered, the wakefield generation is suppressed [see subplot (c)] due to the dispersive effects which become  significant for  $R_0>0.01$. Here, as the value of $R_0$ increases, the magnitude of $1-\delta$ decreases, and so, from Eqs. \eqref{reduced-electron-main-eq} and \eqref{reduced-cpem-main-eq}, the contributions  from the  nonlinear terms  $\propto (1-\delta)$ becomes lower compared to the dispersive terms   $\propto (v_g^2-\delta)$ and $(1-v_g^2)$. Thus, it follows that a transition from dispersive to non-dispersive pulses (wakefields) occurs when the group velocity of the EM wave is close to $c$ and  electrons are  weakly relativistic degenerate \cite{amol2018}. Such a lower value of the degeneracy parameter $R_0$ or the number density $n_0$ and higher value of $v_g~(\sim c)$ is a consequence to the fact that $v_g$ depends on the plasma number density \cite{amol2018}, i.e.,  $d\omega/dk\equiv v_g=k/\omega=\sqrt{1-1/\omega^2}$, i.e., in dimensional form, $v_g=c\sqrt{1-\omega_p^2/\omega^2}$, which can be obtained from the linear dispersion relation of Eq. \eqref{cpem-main-eq}.  An estimate of the amplitude $(a_0)$ dependence of $v_g$  can also be obtained by replacing $n_0$ by $n_{l0}/\gamma_0$ as \cite{amol2018}   $v_g=c\sqrt{1-n_{l0}/n_c\gamma_0}$, where $\gamma_0=\sqrt{1+a_0^2}$ and $n_c=m\omega^2/4\pi e^2$ is the critical number density above which   a transition from the wakefield generation to the soliton formation can take place.  Thus,  for a fixed value of $a_0$, higher values of  $v_g$ close to $c$ correspond  to the lower density regimes (i.e., far below the critical density $n_c$).    As one goes from the regimes of weak  to strong relativistic degenerate plasmas by increasing  the values of $R_0$ or the plasma number density,   the dispersive effect dominates over the nonlinearity. The amplitude of the wakefield  gradually decreases, and it tends to vanish  for sufficiently large values of the degeneracy parameter $R_0$. 
  \subsection{Transition from wakefield generation to soliton formation}\label{sec-sub-transition}
\begin{figure*}
\includegraphics[scale=0.5]{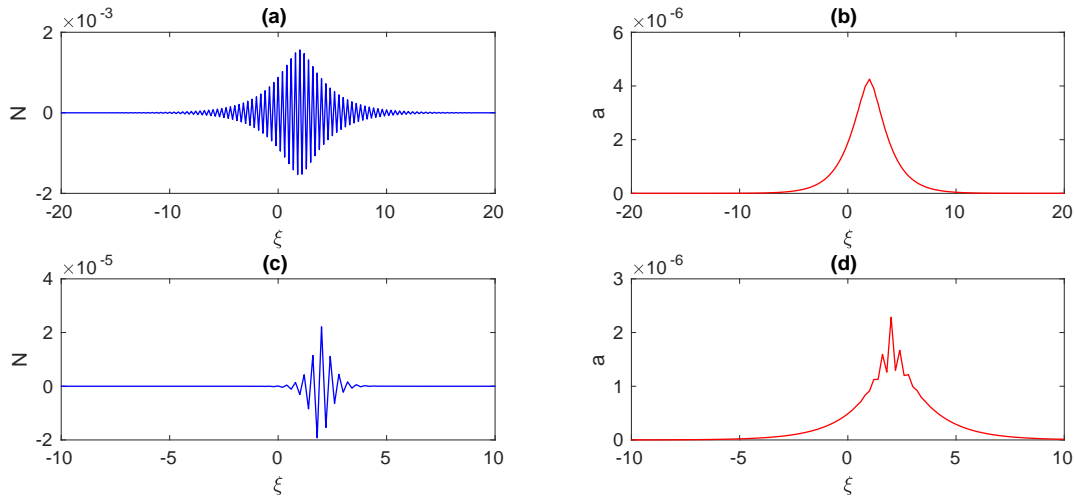}
\caption{ \label{fig2} The transition from the wakefield generation to the soliton formation is shown  with the same parameter values as in Fig. \ref{fig2} but with a lower value of $v_g\sim0.1$  }. 
\end{figure*}

 We note that by retaining the degeneracy parameter in the domain  $0\lesssim R_0 \lesssim 0.01$ and the EM wave frequency in $0.67\lesssim\omega<1$ with  a fixed pulse size $L_p=0.1$ [\textit{cf}. Fig. \ref{fig1}], as the value of the group velocity $v_g$ is reduced from $0.9$ to $0.1$,  the dispersive terms in Eqs. \eqref{reduced-electron-main-eq} and \eqref{reduced-cpem-main-eq} again become dominant over the nonlinearities, and eventually a soliton-like structure forms in the plasma (Fig. \ref{fig2}), i.e., the wakefield generation is no longer possible.      Now, if we gradually enter from the regime of weak to strong relativistic degenerate plasmas by increasing the value of $R_0$ (lower panel of Fig. \ref{fig2}), the number of oscillations decrease together with a reduction of the wave amplitudes. Since we have noticed that the generation of wakefield also occurs in some other parameter regimes with $0.01\lesssim R_0\lesssim0.014,~v_g=0.43,~L_p=0.1$, and $ \omega=0.82$, reducing the value of $v_g$ from $0.43$ to $0.1$, keeping all others unchanged, can also lead to the localization of longitudinal waves similar to Fig. \ref{fig2}. 
 \subsection{Generation of EM soliton}\label{sec-sub-em-soliton}
 In this subsection, we consider the generation of EM solitons by solving Eqs.   \eqref{reduced-electron-main-eq} and \eqref{reduced-cpem-main-eq} with higher values of the wave amplitude $a_0$ and the degeneracy parameter $R_0$.   Before proceeding to the numerical solutions, we first look for some analytic solution of these equations in some particular cases of interest. Thus,  linearizing Eqs. \eqref{reduced-electron-main-eq} and \eqref{reduced-cpem-main-eq} in the limits of $N\ll1$ and $a\ll1$, we obtain
 \begin{equation}
 \left(v_g^2-\delta\right)\frac{d^2N}{d\xi^2}+N=0. \label{linearized-electron-main-eq}
 \end{equation}
 \begin{equation}
  \frac{d^2a}{d\xi^2}+a\omega^2=\frac{a}{1-v_g^2}. \label{linearized-cpem-main-eq}
 \end{equation}
 Then, looking for a solution of the form $\sim\exp(\lambda\xi)$, it is found that Eq. \eqref{linearized-electron-main-eq} has either two real or two purely imaginary eigenvalues $\lambda_{N}^2=1/(\delta-v_g^2)$ according to when $v_g\lessgtr\sqrt{\delta}$, and Eq. \eqref{linearized-cpem-main-eq} has either two real or two imaginary eigenvalues $\lambda_{a}^2=1/(1-v_g^2)-\omega^2$ depending on whether $\omega^2$ is smaller or larger than $1/(1-v_g^2)$, respectively. Thus, EM soliton solution can be found   for $ 1-1/ \omega^2<v_g^2<\delta<1$.
 \par
Within the quasineutrality approximation, i.e.,  $N\approx 1$,   Eq. \eqref{reduced-cpem-main-eq} reduces to
  \begin{equation}
  \frac{d^2a}{d\xi^2}+\left[\omega^2-\frac{1+\left(1-\delta\right)\left(1-\alpha a^2\right)}{1-v_g^2}\right]a=0, \label{quasi-neutral-cpem-eq}
  \end{equation}
 which has  a single humped pure solitary solution of the form
  \begin{equation}
  \begin{split}  
   a(\xi)= &\sqrt{\frac{2-\delta-\omega^2(1-v_g^2)}{\alpha(1-\delta)}}\\&\times \textrm{ sech} \left[\xi  \sqrt{\frac{2-\delta}{1-v_g^2}-\omega^2} \right], \label{soln-quasi-neutral-cpem-eq}
   \end{split}
  \end{equation}
\begin{figure*}
\includegraphics[scale=0.4]{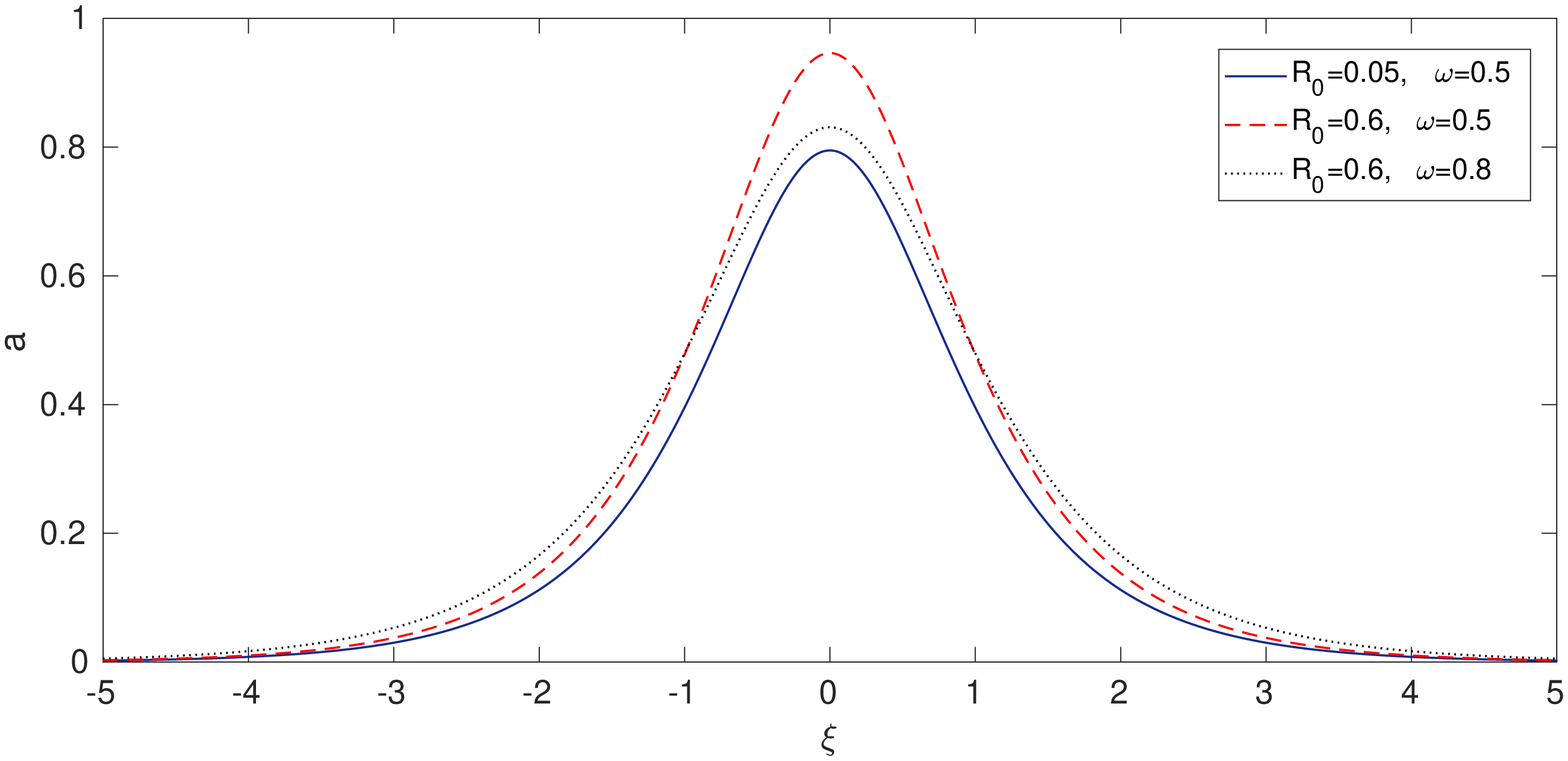}
\caption{\label{fig3} The amplitude of the soliton solution given by Eq \eqref{soln-quasi-neutral-cpem-eq} is shown against the spatial variable $\xi$ for different values of the relativistic degeneracy parameter $ R_0$ and the EM soliton frequency $\omega$,  and for a fixed $v_g=1.1$. It is seen that  both the amplitude and width of the soliton increase  with increasing values of $R_0$. However, for increasing values of $\omega$, the amplitude decreases but the width of the soliton increases. } 
\end{figure*}
We note that both the amplitude and width  of the single hump EM soliton, given by Eq. \eqref{soln-quasi-neutral-cpem-eq}, depend on the degeneracy parameter $R_0$ through $\delta$ and $\alpha$, the soliton frequency $\omega$ and the velocity $v_g$.   It is seen from Fig. \ref{fig3}  that as $R_0$ increases, i.e., as one goes from the regimes of weak  to strong relativistic degenerate plasmas (by increasing the number density), both the amplitude and width of the soliton $a$ increases (see the solid  lines). However, the amplitude decreases, but the width increases with increasing values of the EMW frequency $\omega$. It is also observed that the amplitude of the soliton attains its minimal value   in the regimes of weakly relativistic degenerate plasmas with $R_0\ll1$. 
One can also look for pure EM solitons by increasing the pulse amplitude $a_0=0.3$ and the degeneracy parameter $R_0=0.57$, however,  reducing the group velocity $v_g=0.2$, keeping all other parameters unchanged, i,e.,  $\omega=0.78$ and $L_p=0.1$, so that a delicate balance between the nonlinearity and dispersion takes place. Figure  \ref{fig4} shows the profiles of the EM solitons and relativistic electron density perturbation for different values of $a_0$, $R_0$ and $\omega$.  We note that single hump solitons exist at higher values of the pulse amplitude $a_0$ and the degeneracy parameter $R_0$ [subplots (b) and (c)], and the electron density depletion (hump) is associated with the (hump) dip shaped profiles of the EM solitons. It is also seen that while the amplitudes of the profiles increase  with increasing values of $a_0$ and $\omega$,  those decrease with increasing values of $R_0$. 
\begin{figure*}
\includegraphics[scale=0.5]{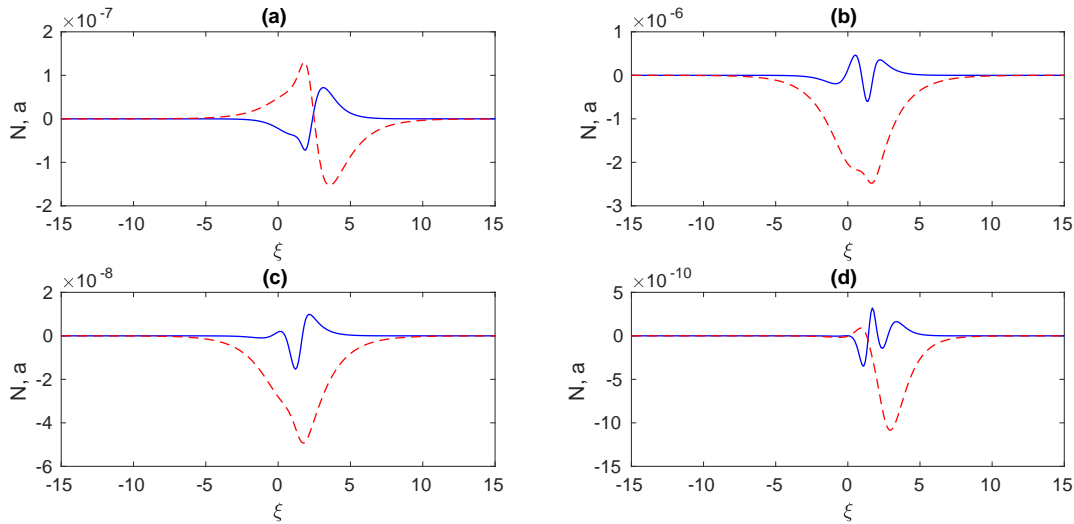}
\caption{\label{fig4} The   generation of EM solitons [numerical solutions of Eqs. \eqref{reduced-electron-main-eq} and \eqref{reduced-cpem-main-eq}] is shown for different values of $a_0,~R_0,~\omega$ and $v_g$: (a)  $a_0=0.3,~R_0=0.57,~\omega=0.78,~v_g=0.2$ and $L_p=0.1$;   (b)  $a_0=0.5,~R_0=0.57,~\omega=0.78,~v_g=0.2$ and $L_p=0.1$;  (c)  $a_0=0.5,~R_0=0.6,~\omega=0.78,~v_g=0.2$ and $L_p=0.1$; (d)  $a_0=0.5,~R_0=0.57,~\omega=0.4,~v_g=0.2$ and $L_p=0.1$.  The  solid and dashed lines correspond to the soliton solutions for  $N$ and $a$ respectively.} 
\end{figure*}
 \par
\section{\label{sec-conclusion} Conclusion}
Starting from a  relativistic fluid model \cite{misra2018} and reducing those into a set of coupled equations for EM wave field and electrostatic density perturbations, we have investigated numerically the generation of wakefields and  their transition into the formation of   soliton-like structures, as well the formation of EM solitons in degenerate dense plasmas.   
Numerical simulation reveals that the generation of wakefields by low-amplitude $(a_0\sim0.01)$ EM pulses is possible in relatively low-density plasmas, i.e.,   when   the degeneracy parameter $R_0$ is in the domain $0<R_0\lesssim0.01$ and the group velocity of EM waves approaches the velocity of light in vacuum $c$, i.e., $v_g\sim0.9$, and  for a fixed pulse size $L_p\sim0.1$ and frequency in the regime $0.67\lesssim\omega<1$. Such  wakefileds can also be generated in relatively high-density regimes where $0.01\lesssim R_0\lesssim0.014$, but $v_g$ is relatively low, i.e., $v_g\sim0.43$ and with a fixed $L_p\sim0.1$ and $\omega=0.82$.  If the plasma number density is  relatively high by increasing the degeneracy parameter, say $R_0\sim0.04$,   the wakefield generation can be suppressed    unless the pulse amplitude is further increased, i.e., $a_0\sim0.3$. Meanwhile retaining the degeneracy parameter $R_0$ in the regime $0<R_0\lesssim0.01$ together with $0.67\lesssim\omega<1$ and $L_p=0.1$, if we decrease $v_g$ from $0.9$ to $0.1$, the plasma density perturbations get localized with a train of oscillations. The number of oscillations get reduced with an increased value of $R_0$.  Such a reduction of the group velocity for the localization of soliton-like structures corresponds to the density enhancement as evident from the linear dispersion relation.   Furthermore, we have seen that the parameter regimes for which the instability of EM solitons (Fig. \ref{fig2}) can be avoided   are  $0<\omega\lesssim1$, $0<v_g\lesssim0.5$ and $R_0\gtrsim0.05$. However, the detailed discussion on soliton stability is left for a future work.
\par 
We have also studied the formation of EM solitons in relativistic degenerate dense plasmas. When $R_0$ is increased further, say $R_0\sim0.57$ together with the pulse amplitude $a_0\sim0.3$ and $v_g$ remains low at $v_g\sim0.2$, keeping other parameters, namely $\omega=0.78$ and $L_p=0.1$ as fixed, the true   soliton formation eventually occurs for both the longitudinal and transverse perturbations in which the density dip (hump) is found to be correlated with the hump (dip)-shaped EM solitons.  
\par To conclude, the inclusion of ion dynamics in relativistic solitary waves can change the picture drastically. In this case, two types of solitons (high- and low-frequency) may coexist  together with the possibility of soliton breaking \cite{bulanov1999}. The latter can  provide  a new mechanism for both ion and electron accelerations in the nonlinear interaction of high-intensity EM waves with relativistic degenerate plasmas. Though,  to our knowledge,  experiments have not been done, especially in high-density relativistic  degenerate plasmas, the present  laser-plasma interaction model  and the results could  help   next-generation lasers   access unexplored regimes   in  relativistic plasmas and test some exotic predictions of the models. Thus, one can explore   the mysteries of astrophysical objects, including interior of white dwarfs  and magnetars, and uncover the dynamic interaction of inner shell degenerate electrons with highly ionized, heavy nuclei.    Furthermore, the  excitation of wakefields and the formation of solitons may be significantly altered in presence of an external magnetic field \cite{amol2011} in   relativistic degenerate plasmas. However, these are issues for future research works and may be reported elsewhere. 
\section*{Acknowledgments} 
{This work was partially supported by a SERB  (Government of India) sponsored research project with sanction order no. CRG/2018/004475 and  UGC-SAP (DRS, Phase III) programme with sanction order no. F.510/3/DRS-III/2015(SAP-I).}


\section*{References}
\begin{thebibliography}{99}
\bibitem{tajima1979}  Tajima T  and  Dawson J  M 1979 \textit{Phys. Rev. Lett.} \textbf{43}  267.
\bibitem{gorbunov1987}  Gorbunov L  M  and  Kirsanov V  I 1987  \textit{Zh. Eksp. Teor. Fiz.} \textbf{93}  509  [1987 \textit{Sov. Phys. JETP} \textbf{66}  290].
\bibitem{balakirev2000}  Balakirev V  A,  Karas I  V  and  Sotnikov G  V 2000 \textit{Phys. Reports}  \textbf{26}  889.
\bibitem{malka2008}   Malka V \textit{et al} 2008 \textit{Nature Phys.} \textbf{4}, 447.
\bibitem{martins2010}  Martins S  F \textit{et al} 2010 \textit{ Nature Phys.} \textbf{6}  311.
\bibitem{lu2007}  Lu W \textit{et al} 2007   \textit{ Phys. Rev. ST Accel. Beams} \textbf{10}  061301.
\bibitem{mironov1992}  Mironov V  A \textit{et al} 1992  \textit{Phys. Rev. A} \textbf{46}  R6178.  
\bibitem{chen2002}  Chen P,  Tajima T  and  Takahashi Y 2002 \textit{ Phys. Rev. Lett.} \textbf{89}  161101 .
\bibitem{chen2009}  Chen P \textit{et al} 2009 \textit{ Plasma Phys. Control. Fusion} \textbf{51}  024012.
\bibitem{misra2010}  Misra A P \textit{et al} 2010 \textit{ Phys. Plasmas} \textbf{17} 122306.
\bibitem{brodin1998}   Brodin G  and   Lundberg J 1998  \textit{Phys. Rev. E} \textbf{57}  7041.
\bibitem{shukla2009}  Shukla P K \textit{et al} 2000 \textit{Phys. Lett. A} \textbf{373}  3165.
\bibitem{shukla1999}  Shukla P  K 1999 \textit{Phys. Plasmas} \textbf{6}  1363.
\bibitem{kneip2009}  Kneip S  \textit{et al} 2009  \textit{ Phys. Rev. Lett.} \textbf{103}  035002.
\bibitem{kim2013}  Kim H  T \textit{et al} 2013  \textit{ Phys. Rev. Lett. } \textbf{111}  165002.
\bibitem{leemans2014}  Leemans W  P \textit{et al} 2014  \textit{ Phys. Rev. Lett.} \textbf{113}  245002.
\bibitem{shen2007}  Shen B \textit{et al} 2007 \textit{Phys. Rev. E} \textbf{76}  055402.
\bibitem{najmudin2003}  Najmudin Z  K \textit{et al} 2003  \textit{Phys. Plasmas } \textbf{10}  2017.
\bibitem{trines2009}  Trines R M G M 2009 \textit{Phys. Rev. E} \textbf{79}  056406.
\bibitem{shukla2009}   Shukla P K 2009 \textit{Plasma Phys. Control. Fusion } {\bf 51}  024013.
\bibitem{joshi2018}   Joshi C \textit{et al} 2018 \textit{Plasma Phys. Control. Fusion} {\bf 60}  034001.
\bibitem{litos2016}   Litos M \textit{ et al} 2016  \textit{ Plasma Phys. Control. Fusion} {\bf 58}  034017.
\bibitem{bulanov1999}   Bulanov S V \textit{et al} 1999   \textit{Phys. Rev. Lett.} \textbf{82}  3440.
\bibitem{amol2018}  Holkundkar A R  and   Brodin G 2018  \textit{Phys. Rev. E} \textbf{97}  043204. 
\bibitem{mima1986}   Mima K \textit{et al} 1986 \textit{Phys. Rev. Lett.} \textbf{57}  1421.
\bibitem{kaw1992}   Kaw P K,   Sen A  and   Katsouleas T 1992 \textit{Phys. Rev. Lett.} \textbf{68}  3172.
\bibitem{mikaberidze2015}   Mikaberidze G and   Berezhiani V I 2015 \textit{Phys. Lett. A} \textbf{379}  2730.
\bibitem{esirkepov1998}  Esirkepov T Zh \textit{et al} 1998 \textit{JETP Lett.} \textbf{68}  36.
\bibitem{bulanov1992}   Bulanov S V \textit{et al} 1992 \textit{Phys. Fluids B} \textbf{4}  1935.
\bibitem{shukla2005}   Shukla P K and   Eliasson B 2005 \textit{Phys. Rev. Lett.} \textbf{94}  065002.
\bibitem{misra2010a}   Misra A P \textit{et al} 2010 \textit{Phys. Plasmas} \textbf{82}  056406.
\bibitem{saxena2013}   Saxena V \textit{et al} 2013 \textit{Phys. Lett. A} \textbf{377}  473.
\bibitem{siminos2014}   Siminos E \textit{et al} 2014 \textit{Phys. Rev. E} \textbf{90}  063104.
\bibitem{mamun2012}   Mamun A A, Roy N and Shukla P K 2012 \textit{J. Plasma Phys.} \textbf{78} 683.
\bibitem{roy2012}  Roy N and  Mamun A A  2012 \textit{Phys. Plasmas } \textbf{19} 033705.
\bibitem{misra2018}   Misra A P  and   Chatterjee D 2018 \textit{Phys. Plasmas} \textbf{25}  062116.
\bibitem{chandrasekhar1935}   Chandrasekhar S 1935 \textit{Mon. Not. R. Astron. Soc.} \textbf{95}  207.
\bibitem{amol2011}  Holkundkar A R,    Brodin G  and   Marklund M 2011 \textit{Phys. Rev. E} \textbf{84} 036409.















\end{thebibliography}
\end{document}